\documentclass[11pt]{article}
\usepackage[margin=1in]{geometry}
\PassOptionsToPackage{table}{xcolor}
\usepackage{graphicx}
\usepackage{hyperref}
\usepackage{amsmath}
\usepackage{amsthm}
\usepackage{amsfonts}
\usepackage{color}
\usepackage{multirow}
\usepackage{tabularx,booktabs}
\usepackage{xspace} 
\usepackage{caption}
\usepackage{subcaption}
\usepackage{enumitem}
\usepackage[table]{xcolor}
\usepackage{tcolorbox}
\usepackage{xcolor}
\usepackage{balance}
\usepackage{microtype}

\AtBeginDocument{%
  \providecommand\BibTeX{{%
    \normalfont B\kern-0.5em{\scshape i\kern-0.25em b}\kern-0.8em\TeX}}}

\newcommand{\todo}[1]{}
\renewcommand{\todo}[1]{{\color{red}{#1}}}

\title{Beyond Chunk-Then-Embed: A Comprehensive Taxonomy and Evaluation of Document Chunking Strategies for Information Retrieval}

\author{
Yongjie Zhou\textsuperscript{1*} \qquad
Shuai Wang\textsuperscript{1*\dag} \qquad
Bevan Koopman\textsuperscript{1,2} \qquad
Guido Zuccon\textsuperscript{1,3}
\\[0.5em]
\small\textsuperscript{1}The University of Queensland, Australia \qquad
\textsuperscript{2}CSIRO, Australia \qquad
\textsuperscript{3}Google
\\[0.3em]
\small\texttt{\{yongjie.zhou@student.uq.edu.au, shuai.wang2@uq.edu.au, g.zuccon@uq.edu.au\}}
\\
\small\texttt{Bevan.Koopman@csiro.au}
\\[0.3em]
\small\textsuperscript{*}Equal contribution \qquad \textsuperscript{\dag}Corresponding author
}

\begin{document}

\maketitle

\begin{abstract}
Document chunking is a critical preprocessing step in dense retrieval systems, yet the design space of chunking strategies remains poorly understood. Recent research has proposed several concurrent approaches, including LLM-guided methods (e.g., DenseX and LumberChunker) and contextualized strategies(e.g., Late Chunking), which generate embeddings before segmentation to preserve contextual information. However, these methods emerged independently and were evaluated on benchmarks with minimal overlap, making direct comparisons difficult.

This paper reproduces prior studies in document chunking and presents a systematic framework that unifies existing strategies along two key dimensions: (1) \textbf{segmentation methods}, including structure-based methods (fixed-size, sentence-based, and paragraph-based) as well as semantically-informed and LLM-guided methods; and (2) \textbf{embedding paradigms}, which determine the timing of chunking relative to embedding (pre-embedding chunking vs. contextualized chunking). Our reproduction evaluates these approaches in two distinct retrieval settings established in previous work: in-document retrieval (needle-in-a-haystack) and in-corpus retrieval (the standard information retrieval task).

Our comprehensive evaluation reveals that optimal chunking strategies are task-dependent: simple structure-based methods outperform LLM-guided alternatives for in-corpus retrieval, while LumberChunker performs best for in-document retrieval. Contextualized chunking improves in-corpus effectiveness but degrades in-document retrieval. 
We also find that chunk size correlates moderately with in-document but weakly with in-corpus effectiveness, suggesting segmentation method differences are not purely driven by chunk size.
Our code and evaluation benchmarks are publicly available at \textit{Anonymoused for review}.

\end{abstract}



\section{Introduction}

Dense retrieval systems represent queries and documents as low-dimensional vectors and power applications from web search engines to retrieval-augmented generation (RAG) frameworks~\cite{li2024survey,zuccon2025r2llms,rau2025context}. However, embedding models have fixed-size input windows, requiring long documents to be partitioned into smaller segments: a process known as \textit{document chunking}. The choice of chunking methods significantly impacts retrieval effectiveness: chunks that are too small may lack sufficient context to match relevant queries, while oversized chunks can dilute relevant information with irrelevant content, and chunks that divide up cohesive relevant content might be highly detrimental to downstream tasks.

Document chunking has evolved rapidly along two dimensions: segmentation methods and embedding-chunking ordering. Segmentation methods determine how documents are divided into chunks, while embedding paradigms determine when chunking occurs relative to embedding (e.g., chunking before or after encoding). For segmentation, structure-based methods use fixed windows or linguistic boundaries like paragraphs~\cite{karpukhin2020dense}, while recent work has introduced semantically-informed methods such as proposition-based chunking like DenseX~\cite{chen2024dense} and LLM-guided methods like LumberChunker~\cite{duarte2024lumberchunker}, which leverage large language models to identify discourse-aware boundaries. Embedding-chunking ordering can be done in two ways: On one hand, the traditional \textit{pre-embedding chunking} approach segments first and embeds each chunk independently. On the other hand, \textit{post-embedding chunking} (also called \textit{contextualized chunking})~\cite{gunther2024late} embeds entire documents first using long-context embedding models (thus preserving global context within individual chunk representations) before pooling token embeddings to form a final embedding. 

Despite the aforementioned research, progress remains fragmented as these methods were evaluated in isolation. For instance, while LumberChunker~\cite{duarte2024lumberchunker} demonstrated higher retrieval effectiveness compared to proposition-based chunking (DenseX~\cite{chen2024dense}) and traditional fixed-size segmentation, it was evaluated exclusively on in-document retrieval (locating relevant information within a single long document) using a single commercial embedding model (OpenAI's text-embedding-ada-002). This evaluation neglected both in-corpus retrieval (finding relevant documents within a collection) and the potential of contextualized chunking. Conversely, \textit{post-embedding chunking} via Late Chunking~\cite{gunther2024late} established the benefits of contextualized chunking but restricted its analysis to simple segmentation methods (e.g., fixed-size, sentence-based) on standard in-corpus benchmarks (BEIR), leaving the interplay between advanced segmentation and contextualized chunking unexplored. This lack of unification makes it difficult to determine whether reported gains stem from the chunking strategies themselves or from specific experimental variables, such as dataset composition or embedding model choice.

To address these gaps, we conduct a reproducibility study that unifies prior work under a common evaluation framework, following the tradition of reproducibility efforts in dense retrieval~\cite{wang2023balanced}. We begin by reproducing the core findings of LumberChunker~\cite{duarte2024lumberchunker} and Late Chunking~\cite{gunther2024late} to validate our implementation and establish baselines. Building on this foundation, we investigate the following research questions:

\textbf{RQ1} \textit{How do different segmentation methods compare for standard in-corpus retrieval?} We evaluate fixed-size, sentence, paragraph, semantic, and LLM-guided segmentation methods using pre-embedding chunking on standard BEIR benchmarks.

\textbf{RQ2} \textit{Do segmentation methods generalize across retrieval tasks?} We apply the same segmentation methods to in-document retrieval, testing whether the relative strengths observed for in-corpus retrieval hold when retrieval is constrained to individual documents.

\textbf{RQ3} \textit{Does contextualized chunking provide consistent improvements across configurations?} We apply contextualized chunking to all segmentation methods from RQ1 across both retrieval tasks to determine whether it provides universal benefits.

\textbf{RQ4} \textit{How does chunk size impact retrieval effectiveness?} Different segmentation methods produce chunks of varying sizes, yet prior work has not controlled for this variable. This raises a simple question: do effectiveness differences reflect segmentation quality, or merely chunk size? At one extreme, treating an entire document as a single chunk guarantees retrieval for in-document settings (as relevant information definitely exists in the chunk), artificially inflating effectiveness. At the other extreme, very large chunks reduce the distinction between contextualized chunking and pre-embedding chunking, as less surrounding context remains to be incorporated; this is particularly true when chunk size approaches the model's context window. We investigate chunk size effects to disentangle these artifacts from genuine segmentation quality.

Through these research questions, we (1) confirm the reproducibility of prior findings (Sec~\ref{sec:reproducibility}), (2) establish baselines for segmentation methods for in-corpus retrieval (Sec~\ref{sec:rq1}), (3) identify which segmentation methods generalize across retrieval tasks (Sec~\ref{sec:rq2}), (4) assess whether contextualized chunking provides universal improvements (Sec~\ref{sec:rq3}), and (5) investigate how much of the observed improvement is attributable to chunk size (Sec~\ref{sec:rq4}).
\section{Document Chunking Strategies}
\label{sec:benchmarking_document_segmentation_strategies}

\begin{figure*}[t]
	\centering
	\includegraphics[width=\textwidth]{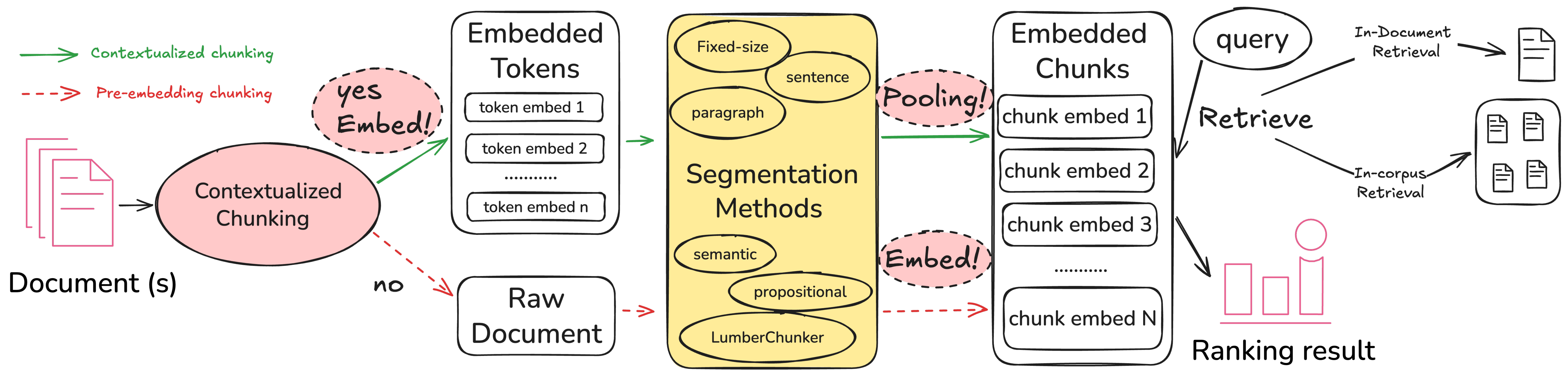}
	\caption{Unified framework for evaluating document chunking strategies across retrieval tasks.}
	\label{fig:chunking-embedding-pipeline}
\end{figure*}

Document chunking transforms raw text into the retrievable chunks used by dense retrieval systems. To systematically compare the wide range of available methods, we organize them along two dimensions: (1) \textbf{segmentation method}, which defines how textual boundaries are identified, and (2) \textbf{embedding-chunking ordering}, which determines whether chunking occurs before or after embedding. This framework provides a structured basis for our experiments and clarifies the relationships between different techniques. Figure~\ref{fig:chunking-embedding-pipeline} gives an overview of the different techniques that we cover in this section.

\subsection{Taxonomy of Segmentation Methods}
We classify segmentation methods into two families: structure-based methods that rely on the document's explicit layout, and semantically-informed methods that leverage language understanding to identify logical divisions. Our taxonomy combines methods from both the LumberChunker and Late Chunking studies~\cite{duarte2024lumberchunker, gunther2024late}.\footnote{We exclude Recursive Chunking and HyDE from the LumberChunker paper as these methods lack sufficient implementation details for reproduction.}

\subsubsection{Structure-Based Methods}
These methods use observable textual features to define chunk boundaries. They are widely adopted in practice due to their computational efficiency, simplicity, and deterministic nature. We evaluate three common methods:

\begin{itemize}[leftmargin=*,nosep]
	\item \textbf{Paragraph:} This method partitions documents at newline characters and removes empty chunks. It respects the author's intended structure at no additional processing cost and is often effective for well-organized texts, such as news articles.
	\item \textbf{Fixed-size:} This method divides text into chunks of a constant token length (e.g., 512 tokens), typically without overlap. While highly efficient, it can arbitrarily split coherent sentences or concepts, potentially degrading embedding quality~\cite{karpukhin2020dense}.
	\item \textbf{Sentence:} This method employs a regex-based splitter to identify sentence boundaries, often grouping multiple sentences per chunk. This creates chunks that balance granularity with context preservation, suitable for tasks where answers may span several sentences.
\end{itemize}

\subsubsection{Semantic and LLM-Guided Methods}
These methods employ models to analyze textual meaning and identify more natural breakpoints. The objective is to produce internally coherent chunks that better support retrieval.

\begin{itemize}[leftmargin=*,nosep]
	\item \textbf{Semantic:} This method calculates the semantic similarity between adjacent sentences using embeddings. A boundary is inserted when the similarity falls below a predefined threshold, indicating a thematic shift~\cite{kamradt5levels}. This allows the segmentation to adapt to the document's content flow.
	\item \textbf{Proposition:} This method uses an LLM to decompose text into atomic facts or ``propositions''~\cite{chen2024dense}. Each proposition serves as a standalone chunk, creating highly specific units suited for factual retrieval. However, this approach significantly increases the total number of chunks and incurs high computational costs due to LLM inference.
	\item \textbf{LumberChunker:} This method directly prompts an LLM to segment a document by inserting breakpoints between paragraphs where it detects a topic shift~\cite{duarte2024lumberchunker}. While effective at capturing discourse structure, it is constrained by the latency and cost associated with LLM inference.
\end{itemize}
Notably, semantically-informed and LLM-guided methods often build upon structured methods. For instance, semantic splitting requires an initial sentence-level pass, while both proposition-based and LumberChunker typically operate on pre-identified paragraphs.

\subsection{Embedding-Chunking Ordering}

The second dimension of our framework is the \textbf{embedding-chunking ordering}, which defines the sequence of chunking and embedding operations. The \textbf{pre-embedding chunking} paradigm, the most common approach, segments the document first and then embeds each resulting chunk in isolation. While simple and scalable, this approach ignores contextual information that may exist across chunk boundaries.

Conversely, \textbf{contextualized chunking} (also known as late chunking) reverses this order. A long-context embedding model first processes the entire document to generate token-level embeddings that incorporate global context~\cite{gunther2024late,rau2025context}. Segmentation boundaries are then applied to these context-aware representations. The final embedding for each chunk is produced by pooling (e.g., averaging) its token embeddings. This approach preserves cross-chunk relationships, though the added context may reduce discriminability between chunks within the same document.

\subsection{Retrieval Tasks}

To address the issue in comparability found when examining prior work, we evaluate chunking strategies across two retrieval tasks:

\begin{itemize}[leftmargin=*,nosep]
	\item \textbf{In-document retrieval:} Often referred to as ``needle-in-a-hay\\stack,'' this task requires locating a specific piece of information within a single long document. It evaluates a method's ability to identify relevant content within a narrow scope.
	\item \textbf{In-corpus retrieval:} This task requires finding the most relevant documents within a large collection, such as the BEIR benchmarks~\cite{thakur2021beir}. It evaluates effectiveness when retrieval spans multiple documents across diverse domains.
\end{itemize}

By evaluating each method across both tasks using a consistent set of embedding models, our study isolates the impact of the chunking strategy itself on retrieval effectiveness.
\section{Experimental Setup}

We evaluate each chunking strategy using the framework described in Section~\ref{sec:benchmarking_document_segmentation_strategies}. Next we detail the specific implementations, datasets, and evaluation procedures employed in our experiments.

\subsection{Segmentation Methods}
We employ a consistent implementation for each segmentation method across all datasets.

For \textbf{structure-based methods}, paragraph-based segmentation partitions documents at newline characters and removes empty chunks. Fixed-size segmentation divides documents into uniform chunks of 256 tokens without overlap. Sentence-based segmentation uses a regular expression-based splitter to identify sentence boundaries and groups 5 sentences per chunk by default. All structure-based methods follow the default protocol and hyperparameters established in Late Chunking~\cite{gunther2024late}.

For \textbf{semantic and LLM-guided methods}, semantic splitting computes the cosine similarity between the embeddings of adjacent sentences and inserts segmentation boundaries where the similarity falls below a specified percentile threshold (95 by default). In our implementation, we use the LlamaIndex semantic node parser\footnote{\texttt{SemanticSplitterNodeParser}} with the \texttt{jina-embeddings-v2-small-en} model, following Late Chunking~\cite{gunther2024late}. Proposition-based segmentation uses an LLM to decompose text into atomic factual units (prompting details are provided in the supplementary materials). LumberChunker analyzes longer text segments to identify optimal breakpoints for each chunk. We employ \texttt{Gemini-2.5-Flash} for both proposition-based segmentation and LumberChunker.\footnote{The original LumberChunker study used \texttt{Gemini-1.0-Pro}. As this model has been deprecated and is no longer accessible via the API, we substituted it with \texttt{Gemini-2.5-Flash}.} For all LLM-guided methods, the generation temperature is set to $T=0.0$ to ensure deterministic and reproducible outputs.

\subsection{Embedding Models}

To assess how different segmentation methods interact with varying embedding architectures, we select four embedding models based on their retrieval effectiveness and architectural compatibility. Since contextualized chunking requires accessing token-level representations prior to pooling, our selection is constrained to high-performing models on the MTEB leaderboard that employ mean pooling. We evaluate the following models:

\begin{itemize}[leftmargin=*,nosep]
	\item \textbf{jina-embeddings-v2-small-en (Jina-v2)}~\cite{gunther2023jina}: A 33M parameter model optimized for efficiency while maintaining competitive effectiveness. This model was a primary baseline in the original Late Chunking study~\cite{gunther2024late}.
	\item \textbf{jina-embeddings-v3 (Jina-v3)}~\cite{gunther2023jina}: A 570M parameter model with improved multilingual capabilities, also central to the evaluation of contextualized chunking.
	\item \textbf{nomic-embed-text-v1 (Nomic)}~\cite{nussbaum2024nomic}: A 137M parameter model with strong effectiveness on diverse benchmarks, used in prior work to evaluate contextualized chunking.
	\item \textbf{multilingual-e5-large (E5-large)}~\cite{wang2024multilingual}: A 560M parameter multilingual model that serves as a standard baseline in retrieval research. We include this model to test whether the observed effects of different chunking strategies generalize beyond the models evaluated in prior work.
\end{itemize}

The first three models were evaluated in the original Late Chunking work~\cite{gunther2024late}; we additionally include E5-large to test generalization across model families.

All models are evaluated under both embedding-chunking orderings. In pre-embedding chunking, documents are segmented first, and each chunk is embedded in isolation. For contextualized (post-embedding) chunking, we feed the maximum sequence of text that fits within the model's context window (e.g., 8,192 tokens). If a document exceeds this limit, we partition it into the minimum necessary number of windows, generating token-level embeddings within the broadest possible context before pooling according to the segmentation boundaries.

\subsection{Datasets and Evaluation}
Our evaluation assesses the generalizability of chunking strategies across two retrieval tasks.

\subsubsection{In-Document Retrieval}

We evaluate in-document retrieval using the \textbf{GutenQA} dataset~\cite{duarte2024lumberchunker}, which comprises 100 classic literary works and 3,000 question-answer pairs. Questions target specific details contained within a single book, evaluating the system's ability to localize information within long documents.

Ground truth annotations in GutenQA are provided at the paragraph level. To accommodate chunks that may span multiple paragraphs or only partially overlap with ground truth, we define a retrieved chunk as relevant if it has any overlap with a ground-truth paragraph.
Following LumberChunker~\cite{duarte2024lumberchunker}, we report Discounted Cumulative Gain at rank 10 (DCG@10), calculated as:
\begin{equation}
	\mathrm{DCG@10} = \sum_{i=1}^{10} \frac{2^{\mathrm{rel}_i} - 1}{\log_2(i+1)}
\end{equation}
where $\mathrm{rel}_i$ represents the relevance score of the result at rank $i$.

\subsubsection{In-Corpus Retrieval}
We evaluate in-corpus retrieval using six datasets from the \textbf{BEIR} benchmark~\cite{thakur2021beir}: FiQA, ArguAna, SciDocs, TREC-COVID, SciFact, and NFCorpus. This extends the four datasets tested in Late Chunking (SciFact, NFCorpus, FiQA, and TREC-COVID). These datasets were selected based on two criteria: first, they cover a diverse range of domains, from financial and scientific text to biomedical literature; second, the corpus sizes are sufficiently compact to make evaluation of computationally intensive LLM-guided methods feasible.

For in-corpus retrieval, relevance is annotated at the document level. Following established practice~\cite{dai2019deeper}, we derive the document-level relevance score from the highest-scoring chunk within that document. This max-pooling (MaxP) strategy is widely used for aggregating chunk-level evidence to produce document rankings.

We report Normalized Discounted Cumulative Gain at rank 10 (nDCG@10) for all BEIR datasets; this metric normalizes DCG@10 against the ideal ranking to produce a value between 0 and 1.
\section{Results}

\subsection{Reproducibility Analysis}
\label{sec:reproducibility}

We begin by verifying the primary claims from the original papers within their original evaluation contexts. This step validates our implementation before extending the analysis to new retrieval tasks.

\begin{table}[htbp]
	\centering
	\caption{Reproduction of LumberChunker evaluation using \texttt{text-embedding-ada-002} on GutenQA (DCG@10).}
	\label{tab:lumber_openai_comparison}
	\begin{tabular}{lcc}
		\toprule
		Segmentation Method & Original & Reproduced \\
		\midrule
		Paragraph & 0.4772 & 0.4879 \\
		Semantic (Jina-v2) & 0.4314 & 0.4457 \\
		Proposition & 0.4565 & 0.4687 \\
		LumberChunker & \textbf{0.6099} & \textbf{0.5953} \\
		\bottomrule
	\end{tabular}
\end{table}


\par\vspace{1em}\noindent\fcolorbox{gray!50}{gray!10}{%
	\parbox{\dimexpr\columnwidth-2\fboxsep-2\fboxrule\relax}{%
		\textbf{Original Claim:} LumberChunker significantly outperforms structure-based methods (e.g., fixed-size, sentence-based) when locating specific information within long documents (in-document retrieval)~\cite{duarte2024lumberchunker}.
	}%
}\par\vspace{1em}

\noindent\textbf{Our Reproduction:} Results in Table~\ref{tab:lumber_openai_comparison} compare segmentation methods using the original embedding model (text-embedding-ada-002) on in-document retrieval. Absolute values show minor differences from the original paper, likely due to the non-deterministic nature of the OpenAI embedding API and our use of Gemini-2.5-Flash instead of the deprecated Gemini-1.0-Pro for LLM-guided segmentation. However, the relative ranking is preserved: LumberChunker achieves the highest effectiveness among all methods, confirming the original findings.

\par\vspace{1em}\noindent\fcolorbox{gray!50}{gray!10}{%
	\parbox{\dimexpr\columnwidth-2\fboxsep-2\fboxrule\relax}{%
		\textbf{Original Claim:} Contextualized chunking (embedding the document before segmentation) preserves global context, thereby improving retrieval effectiveness on standard in-corpus benchmarks compared to embedding chunks independently~\cite{gunther2024late}.
	}%
}\par\vspace{1em}

\noindent\textbf{Our Reproduction:} Results in Table~\ref{tab:late_chunking_comparison} compare contextualized chunking (Con-C) to pre-embedding chunking (Pre-C) using Jina-v2\footnote{All other models from the original study demonstrate the same trend; results are made available in \href{https://anonymous.4open.science/r/Reproduce-chunking-2026-7683}{our repository} due to page limits.}, the main model from the original study. Overall, our reproduction confirms the original findings: contextualized chunking consistently outperforms pre-embedding chunking across datasets and segmentation methods. The only exception is semantic splitting, which shows inconsistent results on FiQA and TREC-COVID. We speculate this discrepancy may stem from version differences in the LlamaIndex SemanticSplitterNodeParser implementation\footnote{The original study did not specify the library version; we used LlamaIndex 0.13.6.}, as this is the only segmentation method relying on an external library with version-dependent behaviour.

\begin{table}[htbp]
	\centering
	\caption{Reproduction of Late Chunking evaluation using Jina-v2 on BEIR datasets (nDCG@10). Pre-C: pre-embedding chunking; Con-C: contextualized chunking. Bold indicates higher value between Pre-C and Con-C.}
	\label{tab:late_chunking_comparison}
	\begin{tabular}{clcccc}
		\toprule
		& & \multicolumn{2}{c}{Original} & \multicolumn{2}{c}{Reproduced} \\
		\cmidrule(lr){3-4} \cmidrule(lr){5-6}
		& Seg. Method & Pre-C & Con-C & Pre-C & Con-C \\
		\midrule
		\multirow{3}{*}{scifact} & Fixed-size & 0.642 & \textbf{0.661} & 0.638 & \textbf{0.652} \\
		& Sentence & 0.647 & \textbf{0.652} & 0.655 & \textbf{0.660} \\
		& Semantic & 0.643 & \textbf{0.650} & 0.645 & \textbf{0.650} \\
		\midrule
		\multirow{3}{*}{nfcorpus} & Fixed-size & 0.235 & \textbf{0.300} & 0.226 & \textbf{0.304} \\
		& Sentence & 0.283 & \textbf{0.300} & 0.285 & \textbf{0.302} \\
		& Semantic & 0.274 & \textbf{0.293} & 0.275 & \textbf{0.293} \\
		\midrule
		\multirow{3}{*}{fiqa} & Fixed-size & 0.333 & \textbf{0.338} & 0.331 & \textbf{0.337} \\
		& Sentence & 0.304 & \textbf{0.339} & 0.310 & \textbf{0.341} \\
		& Semantic & 0.303 & \textbf{0.337} & \textbf{0.303} & 0.223 \\
		\midrule
		\multirow{3}{*}{trec-covid} & Fixed-size & 0.634 & \textbf{0.647} & 0.626 & \textbf{0.647} \\
		& Sentence & 0.665 & \textbf{0.666} & \textbf{0.667} & 0.665 \\
		& Semantic & 0.662 & \textbf{0.663} & \textbf{0.662} & 0.625 \\
		\bottomrule
	\end{tabular}
\end{table}

\noindent\textbf{Summary.} Having confirmed that both LumberChunker's effectiveness advantage for in-document retrieval and Late Chunking's benefits for in-corpus retrieval are reproducible, we proceed to investigate the four research questions. The following sections extend these findings by evaluating all segmentation methods across both retrieval tasks (RQ1--RQ2), examining the interaction between segmentation methods and embedding-chunking ordering (RQ3), and analyzing the confounding effect of chunk size (RQ4).

\subsection{RQ1: Impact of Segmentation Methods on In-Corpus Retrieval}
\label{sec:rq1}

\begin{table*}[htbp]
	\centering
	\caption{Effectiveness under \textbf{pre-embedding chunking (Pre-C)} across models, segmentation methods, and datasets. DCG@10 for GutenQA (in-document); nDCG@10 for BEIR datasets (in-corpus). \textbf{Bold}: best per model-dataset; \underline{underline}: worst. Superscripts: $^a$ = significantly different from best ($p < 0.05$, paired t-test); $^b$ = significantly different from worst.}
	\label{tab:results_regular_encoder}
	\resizebox{\textwidth}{!}{
	\begin{tabular}{llcccccccc}
	\toprule
	&  & \multicolumn{1}{c}{\textbf{\textit{In-document}}} & \multicolumn{7}{c}{\textbf{\textit{In-corpus}}} \\
	\cmidrule(lr){3-3} \cmidrule(lr){4-10}
	& Seg. Method & GutenQA & fiqa & nfcorpus & scifact & trec-covid & arguana & scidocs & Avg \\
	\midrule
	\multirow{6}{*}{\rotatebox{90}{\textbf{Jina-v2}}} & Paragraph & 0.3892$^{ab}$ & \textbf{0.3347}$^{b}$ & 0.3042$^{b}$ & 0.6389 & 0.6518$^{b}$ & 0.4682$^{ab}$ & \textbf{0.1843}$^{b}$ & \textbf{0.4303} \\
	& Sentence & \underline{0.3261}$^{a}$ & 0.3096$^{ab}$ & 0.2847$^{ab}$ & 0.6548 & 0.6674$^{b}$ & 0.4675$^{ab}$ & 0.1732$^{ab}$ & 0.4262 \\
	& Fixed-size & 0.3479$^{ab}$ & 0.3313$^{b}$ & \underline{0.2262}$^{a}$ & \underline{0.6377}$^{a}$ & 0.6256$^{ab}$ & \textbf{0.4803}$^{b}$ & 0.1807$^{ab}$ & 0.4136 \\
	& Semantic & 0.3859$^{ab}$ & 0.3030$^{ab}$ & 0.2746$^{ab}$ & 0.6445 & 0.6618$^{b}$ & 0.4022$^{ab}$ & 0.1760$^{ab}$ & 0.4104 \\
	& LumberChunker & \textbf{0.4659}$^{b}$ & 0.3044$^{ab}$ & 0.2849$^{ab}$ & \textbf{0.6630}$^{b}$ & \textbf{0.6890}$^{b}$ & 0.4472$^{ab}$ & 0.1735$^{ab}$ & 0.4270 \\
	& Proposition & 0.4214$^{ab}$ & \underline{0.2105}$^{a}$ & \textbf{0.3058}$^{b}$ & 0.6410 & \underline{0.4375}$^{a}$ & \underline{0.3589}$^{a}$ & \underline{0.1367}$^{a}$ & \underline{0.3484} \\
	& \textbf{Avg} & 0.3894 & 0.2989 & 0.2801 & 0.6467 & 0.6222 & 0.4374 & 0.1707 & 0.4093 \\
	\midrule
	\multirow{6}{*}{\rotatebox{90}{\textbf{Jina-v3}}} & Paragraph & 0.4574$^{ab}$ & \textbf{0.4747}$^{b}$ & \textbf{0.3663} & \textbf{0.7247}$^{b}$ & \textbf{0.7721}$^{b}$ & 0.4325$^{b}$ & \textbf{0.1984}$^{b}$ & \textbf{0.4948} \\
	& Sentence & \underline{0.3877}$^{a}$ & 0.4333$^{ab}$ & 0.3566 & 0.7163 & 0.7135$^{ab}$ & 0.4256$^{ab}$ & 0.1888$^{ab}$ & 0.4723 \\
	& Fixed-size & 0.4350$^{ab}$ & 0.4680$^{b}$ & 0.3552 & 0.7169 & 0.7385$^{ab}$ & \textbf{0.4362}$^{b}$ & 0.1948$^{ab}$ & 0.4849 \\
	& Semantic & 0.4105$^{ab}$ & 0.4396$^{ab}$ & 0.3596 & 0.7103 & 0.7474$^{b}$ & 0.3894$^{ab}$ & 0.1890$^{ab}$ & 0.4726 \\
	& LumberChunker & \textbf{0.5640}$^{b}$ & 0.4308$^{ab}$ & 0.3590 & 0.7125 & 0.7081$^{ab}$ & 0.4152$^{ab}$ & 0.1881$^{ab}$ & 0.4690 \\
	& Proposition & 0.4182$^{ab}$ & \underline{0.2808}$^{a}$ & \underline{0.3524} & \underline{0.6924}$^{a}$ & \underline{0.5606}$^{a}$ & \underline{0.2931}$^{a}$ & \underline{0.1535}$^{a}$ & \underline{0.3888} \\
	& \textbf{Avg} & 0.4455 & 0.4212 & 0.3582 & 0.7122 & 0.7067 & 0.3987 & 0.1854 & 0.4637 \\
	\midrule
	\multirow{6}{*}{\rotatebox{90}{\textbf{Nomic}}} & Paragraph & 0.4892$^{ab}$ & 0.3802$^{b}$ & \textbf{0.3509} & 0.7047 & 0.7652$^{b}$ & 0.4904$^{ab}$ & 0.1808$^{b}$ & 0.4787 \\
	& Sentence & \underline{0.4146}$^{a}$ & 0.3620$^{ab}$ & 0.3502 & \textbf{0.7150} & 0.7472$^{b}$ & \textbf{0.5084}$^{b}$ & 0.1732$^{ab}$ & 0.4760 \\
	& Fixed-size & 0.4643$^{ab}$ & \textbf{0.3855}$^{b}$ & 0.3484 & 0.7029 & 0.7584$^{b}$ & 0.5042$^{b}$ & \textbf{0.1812}$^{b}$ & \textbf{0.4801} \\
	& Semantic & 0.4274$^{a}$ & 0.3562$^{ab}$ & 0.3506 & 0.7040 & 0.7431$^{b}$ & 0.4434$^{ab}$ & 0.1707$^{ab}$ & 0.4613 \\
	& LumberChunker & \textbf{0.6016}$^{b}$ & 0.3582$^{ab}$ & 0.3502 & 0.7047 & \textbf{0.7762}$^{b}$ & 0.4928$^{ab}$ & 0.1747$^{ab}$ & 0.4762 \\
	& Proposition & 0.4525$^{ab}$ & \underline{0.2554}$^{a}$ & \underline{0.3373} & \underline{0.7025} & \underline{0.5624}$^{a}$ & \underline{0.3660}$^{a}$ & \underline{0.1384}$^{a}$ & \underline{0.3937} \\
	& \textbf{Avg} & 0.4750 & 0.3496 & 0.3479 & 0.7056 & 0.7254 & 0.4675 & 0.1698 & 0.4610 \\
	\midrule
	\multirow{6}{*}{\rotatebox{90}{\textbf{E5-Large}}} & Paragraph & 0.4663$^{ab}$ & \textbf{0.4617}$^{b}$ & 0.3566 & \textbf{0.7129}$^{b}$ & 0.6634 & \textbf{0.5999}$^{b}$ & \textbf{0.2038}$^{b}$ & \textbf{0.4997} \\
	& Sentence & 0.4063$^{ab}$ & 0.4315$^{ab}$ & 0.3583 & 0.6873 & 0.6712$^{b}$ & 0.5646$^{ab}$ & 0.1836$^{ab}$ & 0.4828 \\
	& Fixed-size & 0.4363$^{ab}$ & 0.4509$^{ab}$ & \underline{0.3479} & 0.6740$^{a}$ & 0.6601 & 0.5905$^{ab}$ & 0.1940$^{ab}$ & 0.4863 \\
	& Semantic & \underline{0.3186}$^{a}$ & 0.4173$^{ab}$ & \textbf{0.3583} & 0.6892$^{a}$ & 0.6784$^{b}$ & 0.5540$^{ab}$ & 0.1888$^{ab}$ & 0.4810 \\
	& LumberChunker & \textbf{0.5495}$^{b}$ & 0.4190$^{ab}$ & 0.3557 & 0.6884 & \textbf{0.6952}$^{b}$ & 0.5560$^{ab}$ & 0.1840$^{ab}$ & 0.4830 \\
	& Proposition & 0.4277$^{ab}$ & \underline{0.2539}$^{a}$ & 0.3576 & \underline{0.6705}$^{a}$ & \underline{0.5939}$^{a}$ & \underline{0.3030}$^{a}$ & \underline{0.1509}$^{a}$ & \underline{0.3883} \\
	& \textbf{Avg} & 0.4341 & 0.4057 & 0.3557 & 0.6871 & 0.6604 & 0.5280 & 0.1842 & 0.4702 \\
	\bottomrule
	
	\end{tabular}
	}
\end{table*}
To establish baseline effectiveness of different segmentation methods for in-corpus retrieval, we evaluate six segmentation methods across six BEIR datasets using four embedding models under pre-embedding chunking. Table~\ref{tab:results_regular_encoder} presents results for this analysis.

\par\vspace{1em}\noindent\fcolorbox{gray!50}{gray!10}{%
	\parbox{\dimexpr\columnwidth-2\fboxsep-2\fboxrule\relax}{%
		\textbf{Finding 1:} Structure-based methods outperform semantic/LLM-guided methods, with minimal differences among themselves.
	}%
}\par\vspace{1em}

 Structure-based methods (paragraph, fixed-size, sentence) achieve higher average effectiveness than LLM-guided methods (LumberChunker, proposition-based) and semantic splitting across all four embedding models. Paragraph-based segmentation achieves the highest average nDCG@10 for three models (Jina-v2: 0.4303, Jina-v3: 0.4948, E5-large: 0.4997), while fixed-size performs best with Nomic (0.4801). The effectiveness gap among structure-based methods is small, typically within 1--3\%. For instance, with Jina-v3, paragraph-based (0.4948) outperforms fixed-size by only 2\% (0.4849) and sentence-based by 5\% (0.4723). This suggests that for in-corpus retrieval, uniform and predictable chunk boundaries matter more than identifying semantically ``optimal'' boundaries within each document. For practitioners using pre-embedding chunking, structure-based methods are the clear first choice, and among them, the differences are negligible.

\par\vspace{1em}\noindent\fcolorbox{gray!50}{gray!10}{%
	\parbox{\dimexpr\columnwidth-2\fboxsep-2\fboxrule\relax}{%
		\textbf{Finding 2:} Proposition-based segmentation consistently underperforms all other segmentation methods for in-corpus retrieval.
	}%
}\par\vspace{1em}

The Proposition-based segmentation method yields the lowest average nDCG@10 across all models (ranging from 0.3484 to 0.3937), representing 15--27\% effectiveness degradation compared to paragraph-based segmentation. The degradation is particularly severe on ArguAna (0.29--0.36, representing 35--50\% loss) and FiQA (0.21--0.28, representing 30--40\% loss). This underperformance likely stems from proposition-based segmentation decomposing text into atomic facts, producing very small chunks that lose the topical (contextual) information of the surrounding document. This hypothesis is supported in RQ3, where we show that contextualized chunking, which incorporates surrounding document context into chunk embeddings, substantially improves the effectiveness of proposition-based segmentation.

\par\vspace{1em}\noindent\fcolorbox{gray!50}{gray!10}{%
	\parbox{\dimexpr\columnwidth-2\fboxsep-2\fboxrule\relax}{%
		\textbf{Finding 3:}  The effectiveness of LumberChunker shows high variance across datasets of different domain.
	}%
}\par\vspace{1em}

Unlike proposition-based segmentation, LumberChunker's effectiveness varies substantially across datasets rather than consistently underperforming. It achieves best or near-best results on TREC-COVID (Jina-v2: 0.6890, Nomic: 0.7762, E5-large: 0.6952) but ranks in the bottom half for ArguAna and FiQA. This variance likely reflects LumberChunker's design: it identifies topic shifts within documents, which aligns well with structured scientific abstracts (TREC-COVID) but poorly with argumentative or conversational text where topic boundaries are less defined. For practitioners, LumberChunker may be worth considering if the corpus consists of well-structured documents with clear topic boundaries. However, even in the best case, LumberChunker only matches, not exceeds the effectiveness of structure-based methods, while requiring substantially more computation (Processing Speed: LumberChunker 1.11 docs/s vs. paragraph-based 1,854 docs/s~\footnote{Throughput estimated using SciDocs documents.}). Given that paragraph-based segmentation achieves comparable effectiveness at a fraction of the cost, LumberChunker offers little practical advantage for in-corpus retrieval.

\textbf{Summary.} For in-corpus retrieval with pre-embedding chunking, use structure-based methods. Paragraph-based or fixed-size segmentation are recommended: they are fast, simple, and consistently effective. LLM-guided methods offer no effectiveness advantage despite their computational cost.

\subsection{RQ2: Do Segmentation Methods Generalize Across Retrieval Tasks?}
\label{sec:rq2}
To determine whether the effectiveness rankings observed for in-corpus retrieval generalize to in-document retrieval, we evaluate all six segmentation methods on the GutenQA dataset using the same four embedding models. GutenQA represents a different retrieval challenge: locating relevant passages within individual long-form documents (books) rather than identifying relevant documents within a corpus. Table~\ref{tab:results_regular_encoder} presents results for in-document retrieval (GutenQA column).

\par\vspace{1em}\noindent\fcolorbox{gray!50}{gray!10}{%
	\parbox{\dimexpr\columnwidth-2\fboxsep-2\fboxrule\relax}{%
		\textbf{Finding 1:} Segmentation method effectiveness rankings reverse between in-corpus and in-document retrieval.
	}%
}\par\vspace{1em}

LumberChunker, which ranked 2nd--4th in in-corpus retrieval (RQ1), achieves the highest effectiveness on GutenQA across all four models (Jina-v2: 0.4659, Jina-v3: 0.5640, Nomic: 0.6016, E5-large: 0.5495), outperforming paragraph-based segmentation by 10--30\%. Conversely, paragraph-based segmentation, which performed best in in-corpus retrieval, drops to 2nd--3rd position for in-document retrieval. This confirms our hypothesis from RQ1: LumberChunker's effectiveness depends on document structure. GutenQA consists of literary books with clear narrative structure, which aligns well with LumberChunker's topic-shift detection. In-document retrieval requires distinguishing relevant chunks from other chunks within the same document, and LumberChunker's adaptive boundaries make relevant chunks more distinguishable than uniform paragraph boundaries.

\par\vspace{1em}\noindent\fcolorbox{gray!50}{gray!10}{%
	\parbox{\dimexpr\columnwidth-2\fboxsep-2\fboxrule\relax}{%
		\textbf{Finding 2:} Proposition-based segmentation remains weak but improves relatively.
	}%
}\par\vspace{1em}

Proposition-based segmentation maintains poor effectiveness in both retrieval tasks, but its relative ranking improves from consistently worst in in-corpus retrieval to competitive (ranked second to fourth) on GutenQA. This partial recovery suggests that while atomic facts still lack sufficient context, the in-document retrieval setting is less sensitive to this limitation than cross-document retrieval.

\textbf{Summary.} Segmentation methods do not generalize across retrieval tasks. For in-document retrieval, use LumberChunker; for in-corpus retrieval, use structure-based methods (RQ1). This confirms that LumberChunker's effectiveness depends on document structure: it performs well when documents have clear internal organization (GutenQA's literary works, TREC-COVID's scientific abstracts) but poorly on corpora consisting of less structured documents.

\subsection{RQ3: Effectiveness of Contextualized Chunking}
\label{sec:rq3}
\begin{table*}[htbp]
	\centering
	\definecolor{lightgreen}{RGB}{220,255,220}
	\definecolor{lightred}{RGB}{255,220,220}
	\caption{Relative effectiveness change (\%) from \textbf{pre-embedding chunking (Pre-C)} to \textbf{contextualized chunking (Con-C)}. Green: improvement; Red: degradation. $^*$: statistically significant ($p < 0.05$, paired t-test).}
	\label{tab:encoder_comparison}
	\resizebox{\textwidth}{!}{
	\begin{tabular}{llcccccccc}
\toprule
&  & \multicolumn{1}{c}{\textbf{\textit{In-document}}} & \multicolumn{7}{c}{\textbf{\textit{In-corpus}}} \\
\cmidrule(lr){3-3} \cmidrule(lr){4-10}
& Seg. Method & GutenQA & fiqa & nfcorpus & scifact & trec-covid & arguana & scidocs & Avg \\
\midrule
\multirow{6}{*}{\textbf{Jina-v2}} & Paragraph & \cellcolor{lightred}-10.76$^*$ & \cellcolor{lightred}-0.02 & \cellcolor{lightred}-0.16 & 0.00 & 0.00 & \cellcolor{lightred}-0.21 & \cellcolor{lightgreen}+0.73$^*$ & \cellcolor{lightgreen}+0.06 \\
& Sentence & \cellcolor{lightred}-5.39$^*$ & \cellcolor{lightgreen}+9.97$^*$ & \cellcolor{lightgreen}+6.16$^*$ & \cellcolor{lightgreen}+0.75 & \cellcolor{lightred}-0.37 & \cellcolor{lightgreen}+1.96$^*$ & \cellcolor{lightgreen}+6.48$^*$ & \cellcolor{lightgreen}+4.16 \\
& Fixed-size & \cellcolor{lightred}-2.85$^*$ & \cellcolor{lightgreen}+2.00 & \cellcolor{lightgreen}+34.41$^*$ & \cellcolor{lightgreen}+2.30 & \cellcolor{lightgreen}+3.38 & \cellcolor{lightred}-1.06$^*$ & \cellcolor{lightgreen}+2.18$^*$ & \cellcolor{lightgreen}+7.20 \\
& Semantic & \cellcolor{lightgreen}+4.56$^*$ & \cellcolor{lightred}-26.37$^*$ & \cellcolor{lightgreen}+6.72$^*$ & \cellcolor{lightgreen}+0.83 & \cellcolor{lightred}-5.56 & \cellcolor{lightgreen}+17.49$^*$ & \cellcolor{lightgreen}+3.90$^*$ & \cellcolor{lightred}-0.50 \\
& LumberChunker & \cellcolor{lightred}-10.45$^*$ & \cellcolor{lightgreen}+10.76$^*$ & \cellcolor{lightgreen}+6.75$^*$ & \cellcolor{lightred}-0.62 & \cellcolor{lightred}-1.54 & \cellcolor{lightgreen}+6.40$^*$ & \cellcolor{lightgreen}+6.32$^*$ & \cellcolor{lightgreen}+4.68 \\
& Proposition & \cellcolor{lightred}-6.06$^*$ & \cellcolor{lightgreen}+49.02$^*$ & \cellcolor{lightgreen}+2.95 & \cellcolor{lightgreen}+1.45 & \cellcolor{lightgreen}+26.74$^*$ & \cellcolor{lightgreen}+36.77$^*$ & \cellcolor{lightgreen}+20.27$^*$ & \cellcolor{lightgreen}+22.87 \\
& \textbf{Avg} & \cellcolor{lightred}-5.16 & \cellcolor{lightgreen}+7.56 & \cellcolor{lightgreen}+9.47 & \cellcolor{lightgreen}+0.78 & \cellcolor{lightgreen}+3.78 & \cellcolor{lightgreen}+10.22 & \cellcolor{lightgreen}+6.65 & \cellcolor{lightgreen}+6.41 \\
\midrule
\multirow{6}{*}{\textbf{Jina-v3}} & Paragraph & \cellcolor{lightred}-37.88$^*$ & \cellcolor{lightred}-0.00 & \cellcolor{lightgreen}+0.00 & 0.00 & 0.00 & \cellcolor{lightred}-0.01 & \cellcolor{lightgreen}+0.49 & \cellcolor{lightgreen}+0.08 \\
& Sentence & \cellcolor{lightred}-35.01$^*$ & \cellcolor{lightgreen}+10.66$^*$ & \cellcolor{lightgreen}+3.02 & \cellcolor{lightgreen}+2.46$^*$ & \cellcolor{lightgreen}+7.83$^*$ & \cellcolor{lightred}-1.46 & \cellcolor{lightgreen}+4.90$^*$ & \cellcolor{lightgreen}+4.57 \\
& Fixed-size & \cellcolor{lightred}-35.20$^*$ & \cellcolor{lightgreen}+2.25$^*$ & \cellcolor{lightgreen}+3.67$^*$ & \cellcolor{lightgreen}+1.86 & \cellcolor{lightgreen}+3.66 & \cellcolor{lightred}-2.31$^*$ & \cellcolor{lightgreen}+1.84$^*$ & \cellcolor{lightgreen}+1.83 \\
& Semantic & \cellcolor{lightred}-18.38$^*$ & \cellcolor{lightred}-26.73$^*$ & \cellcolor{lightgreen}+2.05 & \cellcolor{lightgreen}+1.83 & \cellcolor{lightred}-6.53 & \cellcolor{lightgreen}+12.06$^*$ & \cellcolor{lightgreen}+4.93$^*$ & \cellcolor{lightred}-2.07 \\
& LumberChunker & \cellcolor{lightred}-35.19$^*$ & \cellcolor{lightgreen}+9.84$^*$ & \cellcolor{lightgreen}+2.04 & \cellcolor{lightgreen}+1.37 & \cellcolor{lightgreen}+8.06$^*$ & \cellcolor{lightgreen}+1.49 & \cellcolor{lightgreen}+6.02$^*$ & \cellcolor{lightgreen}+4.80 \\
& Proposition & \cellcolor{lightred}-17.25$^*$ & \cellcolor{lightgreen}+54.75$^*$ & \cellcolor{lightgreen}+4.23 & \cellcolor{lightgreen}+3.65$^*$ & \cellcolor{lightgreen}+29.43$^*$ & \cellcolor{lightgreen}+51.04$^*$ & \cellcolor{lightgreen}+19.56$^*$ & \cellcolor{lightgreen}+27.11 \\
& \textbf{Avg} & \cellcolor{lightred}-29.82 & \cellcolor{lightgreen}+8.46 & \cellcolor{lightgreen}+2.50 & \cellcolor{lightgreen}+1.86 & \cellcolor{lightgreen}+7.08 & \cellcolor{lightgreen}+10.13 & \cellcolor{lightgreen}+6.29 & \cellcolor{lightgreen}+6.05 \\
\midrule
\multirow{6}{*}{\textbf{Nomic}} & Paragraph & \cellcolor{lightred}-62.47$^*$ & 0.00 & \cellcolor{lightred}-0.05 & 0.00 & 0.00 & \cellcolor{lightgreen}+0.00 & \cellcolor{lightred}-0.10 & \cellcolor{lightred}-0.03 \\
& Sentence & \cellcolor{lightred}-62.57$^*$ & \cellcolor{lightgreen}+4.95$^*$ & \cellcolor{lightgreen}+1.31 & \cellcolor{lightred}-0.48 & \cellcolor{lightgreen}+4.23$^*$ & \cellcolor{lightred}-4.47$^*$ & \cellcolor{lightgreen}+2.02 & \cellcolor{lightgreen}+1.26 \\
& Fixed-size & \cellcolor{lightred}-63.23$^*$ & \cellcolor{lightgreen}+0.37 & \cellcolor{lightgreen}+0.70 & \cellcolor{lightgreen}+0.56 & \cellcolor{lightred}-1.17 & \cellcolor{lightred}-3.31$^*$ & \cellcolor{lightred}-0.97 & \cellcolor{lightred}-0.64 \\
& Semantic & \cellcolor{lightred}-43.16$^*$ & \cellcolor{lightred}-25.43$^*$ & \cellcolor{lightgreen}+0.81 & \cellcolor{lightgreen}+0.15 & \cellcolor{lightred}-1.81 & \cellcolor{lightgreen}+9.54$^*$ & \cellcolor{lightgreen}+5.05$^*$ & \cellcolor{lightred}-1.95 \\
& LumberChunker & \cellcolor{lightred}-61.66$^*$ & \cellcolor{lightgreen}+6.45$^*$ & \cellcolor{lightgreen}+1.14 & \cellcolor{lightgreen}+1.31 & \cellcolor{lightgreen}+2.46 & \cellcolor{lightred}-1.58 & \cellcolor{lightgreen}+4.75$^*$ & \cellcolor{lightgreen}+2.42 \\
& Proposition & \cellcolor{lightred}-25.94$^*$ & \cellcolor{lightgreen}+29.63$^*$ & \cellcolor{lightgreen}+4.15$^*$ & \cellcolor{lightgreen}+3.90$^*$ & \cellcolor{lightgreen}+9.68$^*$ & \cellcolor{lightgreen}+31.18$^*$ & \cellcolor{lightgreen}+15.16$^*$ & \cellcolor{lightgreen}+15.62 \\
& \textbf{Avg} & \cellcolor{lightred}-53.17 & \cellcolor{lightgreen}+2.66 & \cellcolor{lightgreen}+1.34 & \cellcolor{lightgreen}+0.91 & \cellcolor{lightgreen}+2.23 & \cellcolor{lightgreen}+5.23 & \cellcolor{lightgreen}+4.32 & \cellcolor{lightgreen}+2.78 \\
\midrule
\multirow{6}{*}{\textbf{E5-Large}} & Paragraph & \cellcolor{lightred}-9.74$^*$ & \cellcolor{lightgreen}+0.10 & \cellcolor{lightred}-0.02 & \cellcolor{lightred}-0.16 & \cellcolor{lightgreen}+0.00 & \cellcolor{lightred}-0.05 & \cellcolor{lightgreen}+0.87$^*$ & \cellcolor{lightgreen}+0.12 \\
& Sentence & \cellcolor{lightred}-10.95$^*$ & \cellcolor{lightgreen}+6.78$^*$ & \cellcolor{lightred}-0.94 & \cellcolor{lightgreen}+2.08 & \cellcolor{lightred}-2.84 & \cellcolor{lightgreen}+5.30$^*$ & \cellcolor{lightgreen}+11.82$^*$ & \cellcolor{lightgreen}+3.70 \\
& Fixed-size & \cellcolor{lightgreen}+0.01 & \cellcolor{lightgreen}+2.96$^*$ & \cellcolor{lightgreen}+1.62 & \cellcolor{lightgreen}+6.54$^*$ & \cellcolor{lightgreen}+1.23 & \cellcolor{lightgreen}+0.89 & \cellcolor{lightgreen}+6.20$^*$ & \cellcolor{lightgreen}+3.24 \\
& Semantic & \cellcolor{lightgreen}+1.95$^*$ & \cellcolor{lightred}-23.56$^*$ & \cellcolor{lightred}-1.33 & \cellcolor{lightgreen}+1.28 & \cellcolor{lightred}-7.04 & \cellcolor{lightgreen}+6.99$^*$ & \cellcolor{lightgreen}+8.44$^*$ & \cellcolor{lightred}-2.54 \\
& LumberChunker & \cellcolor{lightred}-3.08$^*$ & \cellcolor{lightgreen}+10.32$^*$ & \cellcolor{lightred}-1.14 & \cellcolor{lightgreen}+2.53 & \cellcolor{lightred}-4.04 & \cellcolor{lightgreen}+6.57$^*$ & \cellcolor{lightgreen}+11.65$^*$ & \cellcolor{lightgreen}+4.32 \\
& Proposition & \cellcolor{lightred}-1.18 & \cellcolor{lightgreen}+62.80$^*$ & \cellcolor{lightred}-2.99 & \cellcolor{lightred}-3.94 & \cellcolor{lightred}-2.18 & \cellcolor{lightgreen}+83.30$^*$ & \cellcolor{lightgreen}+24.65$^*$ & \cellcolor{lightgreen}+26.94 \\
& \textbf{Avg} & \cellcolor{lightred}-3.83 & \cellcolor{lightgreen}+9.90 & \cellcolor{lightred}-0.80 & \cellcolor{lightgreen}+1.39 & \cellcolor{lightred}-2.48 & \cellcolor{lightgreen}+17.17 & \cellcolor{lightgreen}+10.60 & \cellcolor{lightgreen}+5.96 \\
\bottomrule

	\end{tabular}
	}
\end{table*}

To assess whether contextualized chunking provides consistent improvements across segmentation methods and retrieval tasks, we compare it to pre-embedding chunking for all segmentation methods across both in-corpus (BEIR) and in-document (GutenQA) retrieval. Table~\ref{tab:encoder_comparison} presents the relative effectiveness changes (in percentage) when using contextualized chunking compared to pre-embedding chunking, while Table~\ref{tab:results_late_encoder} reports the absolute effectiveness under contextualized chunking.

\par\vspace{1em}\noindent\fcolorbox{gray!50}{gray!10}{%
	\parbox{\dimexpr\columnwidth-2\fboxsep-2\fboxrule\relax}{%
		\textbf{Finding 1:} Contextualized chunking generally improves in-corpus retrieval, with the largest gains for LLM-guided methods.
	}%
}\par\vspace{1em}

As shown in Table~\ref{tab:encoder_comparison}, contextualized chunking improves in-corpus retrieval effectiveness for most segmentation methods. The gains are largest for proposition-based segmentation, with average improvements of +22.87\% (Jina-v2), +27.11\% (Jina-v3), +15.62\% (Nomic), and +26.94\% (E5-large). LumberChunker also shows consistent improvements ranging from +2.42\% to +4.80\% on average. Structure-based methods benefit less, with smaller and sometimes mixed gains. This confirms our hypothesis from RQ1: proposition-based segmentation's poor performance under pre-embedding chunking stems from losing document context, which contextualized chunking restores by incorporating surrounding information into chunk embeddings.

\begin{table*}[htbp]
	\centering
	\caption{Effectiveness under \textbf{contextualized chunking (Con-C)} across models, segmentation methods, and datasets. DCG@10 for GutenQA (in-document); nDCG@10 for BEIR datasets (in-corpus). \textbf{Bold}: best per model-dataset; \underline{underline}: worst. Superscripts: $^a$ = significantly different from best ($p < 0.05$, paired t-test); $^b$ = significantly different from worst.}
	\label{tab:results_late_encoder}
	\resizebox{\textwidth}{!}{
	\begin{tabular}{llcccccccc}
		\toprule
		&  & \multicolumn{1}{c}{\textbf{\textit{In-document}}} & \multicolumn{7}{c}{\textbf{\textit{In-corpus}}} \\
		\cmidrule(lr){3-3} \cmidrule(lr){4-10}
		& Seg. Method  & GutenQA & fiqa & nfcorpus & scifact & trec-covid & arguana & scidocs & Avg \\
		\midrule
		\multirow{6}{*}{\rotatebox{90}{\textbf{Jina-v2}}} & Paragraph & 0.3473$^{ab}$ & 0.3346$^{b}$ & 0.3038$^{ab}$ & \underline{0.6389}$^{a}$ & 0.6518$^{ab}$ & \underline{0.4672}$^{a}$ & \textbf{0.1856}$^{b}$ & 0.4303 \\
		& Sentence & \underline{0.3085}$^{a}$ & \textbf{0.3405}$^{b}$ & 0.3022$^{ab}$ & \textbf{0.6597}$^{b}$ & 0.6649$^{b}$ & 0.4767$^{b}$ & 0.1844$^{b}$ & 0.4381 \\
		& Fixed-size & 0.3380$^{ab}$ & 0.3379$^{b}$ & 0.3040$^{b}$ & 0.6524 & 0.6468$^{ab}$ & 0.4752$^{ab}$ & 0.1846$^{b}$ & 0.4335 \\
		& Semantic & 0.4035$^{ab}$ & \underline{0.2231}$^{a}$ & \underline{0.2931}$^{a}$ & 0.6498 & 0.6250$^{ab}$ & 0.4726$^{a}$ & 0.1829$^{b}$ & \underline{0.4077} \\
		& LumberChunker & \textbf{0.4172}$^{b}$ & 0.3371$^{b}$ & 0.3041$^{b}$ & 0.6589$^{b}$ & \textbf{0.6784}$^{b}$ & 0.4758$^{ab}$ & 0.1844$^{b}$ & \textbf{0.4398} \\
		& Proposition & 0.3959$^{ab}$ & 0.3136$^{ab}$ & \textbf{0.3148}$^{b}$ & 0.6503 & \underline{0.5545}$^{a}$ & \textbf{0.4909}$^{b}$ & \underline{0.1644}$^{a}$ & 0.4148 \\
		& \textbf{Avg} & 0.3684 & 0.3145 & 0.3037 & 0.6517 & 0.6369 & 0.4764 & 0.1811 & 0.4274 \\
		\midrule
		\multirow{6}{*}{\rotatebox{90}{\textbf{Jina-v3}}} & Paragraph & 0.2842$^{ab}$ & 0.4747$^{ab}$ & \underline{0.3663} & 0.7247 & \textbf{0.7721}$^{b}$ & 0.4325$^{b}$ & 0.1994$^{b}$ & \textbf{0.4949} \\
		& Sentence & \underline{0.2520}$^{a}$ & \textbf{0.4795}$^{b}$ & 0.3673 & \textbf{0.7340} & 0.7693$^{b}$ & \underline{0.4194}$^{a}$ & 0.1980$^{b}$ & 0.4946 \\
		& Fixed-size & 0.2818$^{ab}$ & 0.4785$^{b}$ & \textbf{0.3682} & 0.7302 & 0.7656$^{b}$ & 0.4261$^{ab}$ & 0.1984$^{b}$ & 0.4945 \\
		& Semantic & 0.3350$^{ab}$ & \underline{0.3221}$^{a}$ & 0.3670 & 0.7233$^{a}$ & \underline{0.6986}$^{a}$ & 0.4364$^{b}$ & 0.1983$^{b}$ & \underline{0.4576} \\
		& LumberChunker & \textbf{0.3655}$^{b}$ & 0.4732$^{ab}$ & 0.3663 & 0.7223$^{a}$ & 0.7652$^{b}$ & 0.4214$^{a}$ & \textbf{0.1994}$^{b}$ & 0.4913 \\
		& Proposition & 0.3460$^{ab}$ & 0.4345$^{ab}$ & 0.3673 & \underline{0.7176} & 0.7256 & \textbf{0.4426}$^{b}$ & \underline{0.1836}$^{a}$ & 0.4785 \\
		& \textbf{Avg} & 0.3108 & 0.4438 & 0.3671 & 0.7253 & 0.7494 & 0.4297 & 0.1962 & 0.4853 \\
		\midrule
		\multirow{6}{*}{\rotatebox{90}{\textbf{Nomic}}} & Paragraph & 0.1836$^{ab}$ & 0.3802$^{ab}$ & \underline{0.3507} & \underline{0.7047}$^{a}$ & 0.7652$^{b}$ & \textbf{0.4904} & 0.1806$^{b}$ & 0.4786 \\
		& Sentence & \underline{0.1552}$^{a}$ & 0.3799$^{ab}$ & \textbf{0.3548} & 0.7115 & 0.7788$^{b}$ & 0.4857 & 0.1767$^{ab}$ & 0.4812 \\
		& Fixed-size & 0.1708$^{ab}$ & \textbf{0.3869}$^{b}$ & 0.3509 & 0.7069 & 0.7495$^{ab}$ & 0.4875 & 0.1794$^{b}$ & 0.4768 \\
		& Semantic & 0.2429$^{ab}$ & \underline{0.2656}$^{a}$ & 0.3534 & 0.7050$^{a}$ & 0.7296$^{ab}$ & 0.4857 & 0.1793$^{ab}$ & 0.4531 \\
		& LumberChunker & 0.2307$^{ab}$ & 0.3813$^{b}$ & 0.3541 & 0.7140 & \textbf{0.7953}$^{b}$ & 0.4850 & \textbf{0.1830}$^{b}$ & \textbf{0.4855} \\
		& Proposition & \textbf{0.3351}$^{b}$ & 0.3311$^{ab}$ & 0.3513 & \textbf{0.7299}$^{b}$ & \underline{0.6169}$^{a}$ & \underline{0.4801} & \underline{0.1594}$^{a}$ & \underline{0.4448} \\
		& \textbf{Avg} & 0.2197 & 0.3542 & 0.3525 & 0.7120 & 0.7392 & 0.4857 & 0.1764 & 0.4700 \\
		\midrule
		\multirow{6}{*}{\rotatebox{90}{\textbf{E5-Large}}} & Paragraph & 0.4209$^{ab}$ & 0.4622$^{b}$ & \textbf{0.3565} & 0.7118$^{b}$ & 0.6634$^{b}$ & \textbf{0.5996}$^{b}$ & 0.2055$^{b}$ & 0.4998 \\
		& Sentence & 0.3618$^{ab}$ & 0.4607$^{b}$ & 0.3549 & 0.7016$^{ab}$ & 0.6522$^{b}$ & 0.5946$^{ab}$ & 0.2053$^{b}$ & 0.4949 \\
		& Fixed-size & 0.4364$^{ab}$ & \textbf{0.4643}$^{b}$ & 0.3536 & \textbf{0.7181}$^{b}$ & \textbf{0.6682}$^{b}$ & 0.5958$^{ab}$ & \textbf{0.2061}$^{b}$ & \textbf{0.5010} \\
		& Semantic & \underline{0.3248}$^{a}$ & \underline{0.3190}$^{a}$ & 0.3535 & 0.6980$^{ab}$ & 0.6307$^{a}$ & 0.5927$^{ab}$ & 0.2047$^{b}$ & 0.4664 \\
		& LumberChunker & \textbf{0.5326}$^{b}$ & 0.4622$^{b}$ & 0.3517 & 0.7058$^{ab}$ & 0.6671$^{b}$ & 0.5925$^{ab}$ & 0.2055$^{b}$ & 0.4975 \\
		& Proposition & 0.4227$^{ab}$ & 0.4133$^{ab}$ & \underline{0.3469} & \underline{0.6441}$^{a}$ & \underline{0.5810}$^{a}$ & \underline{0.5553}$^{a}$ & \underline{0.1881}$^{a}$ & \underline{0.4548} \\
		& \textbf{Avg} & 0.4165 & 0.4303 & 0.3528 & 0.6966 & 0.6437 & 0.5884 & 0.2025 & 0.4857 \\
		\bottomrule
		
	\end{tabular}
	}
\end{table*}

\par\vspace{1em}\noindent\fcolorbox{gray!50}{gray!10}{%
	\parbox{\dimexpr\columnwidth-2\fboxsep-2\fboxrule\relax}{%
		\textbf{Finding 2:} Large relative gains do not guarantee high absolute effectiveness.
	}%
}\par\vspace{1em}

Despite proposition-based segmentation achieving the largest relative improvements from contextualized chunking, it does not yield the highest absolute effectiveness. As shown in Table~\ref{tab:results_late_encoder}, proposition-based methods still rank below structure-based methods and LumberChunker in absolute in-corpus effectiveness for most configurations. For example, with Jina-v3, proposition-based achieves 0.4785 average nDCG@10 under contextualized chunking, while paragraph-based achieves 0.4949 and LumberChunker achieves 0.4913. The large percentage improvements reflect recovery from a very low pre-embedding baseline rather than achieving top-tier effectiveness. For practitioners, contextualized chunking makes proposition-based segmentation viable but not optimal.

\par\vspace{1em}\noindent\fcolorbox{gray!50}{gray!10}{%
	\parbox{\dimexpr\columnwidth-2\fboxsep-2\fboxrule\relax}{%
		\textbf{Finding 3:} Contextualized chunking degrades in-document retrieval across all configurations.
	}%
}\par\vspace{1em}

While contextualized chunking improves in-corpus retrieval, it consistently degrades in-document retrieval effectiveness on GutenQA across all segmentation methods and models. The severity varies by model: Nomic suffers the most severe losses (averaging $-$53.17\%), followed by Jina-v3 ($-$29.82\%), while Jina-v2 ($-$5.16\%) and E5-large ($-$3.83\%) show more moderate degradation. Recall from RQ2 that in-document retrieval requires distinguishing relevant chunks from other chunks within the same document. We hypothesize that contextualized chunking produces embeddings that encode broader document themes rather than chunk-specific details, making chunks within the same document more similar to each other and thus harder to distinguish.

\par\vspace{1em}\noindent\fcolorbox{gray!50}{gray!10}{%
	\parbox{\dimexpr\columnwidth-2\fboxsep-2\fboxrule\relax}{%
		\textbf{Finding 4:} The effect of contextualized chunking varies by model and dataset.
	}%
}\par\vspace{1em}

The impact of contextualized chunking depends on both the embedding model and the dataset. With respect to models, Jina-v2, Jina-v3, and Nomic show more consistent improvements on in-corpus retrieval, while E5-large shows mixed results with some configurations degrading. This difference likely stems from context window size: Jina-v2, Jina-v3, and Nomic support long context inputs (8,192 tokens), allowing contextualized chunking to incorporate substantial surrounding document information, whereas E5-large has a 512-token context limit, restricting how much additional context can be leveraged. With respect to datasets, FiQA shows notable degradation for semantic splitting across all models ($-$26.37\% for Jina-v2, $-$26.73\% for Jina-v3, $-$25.43\% for Nomic, $-$23.56\% for E5-large), despite improvements on other datasets. Looking at Table~\ref{tab:results_regular_encoder}, semantic splitting already underperforms structure-based methods on FiQA under pre-embedding chunking, suggesting that semantic boundaries do not align well with this corpus of financial forum discussions. Contextualized chunking makes this worse: incorporating context from poorly-defined boundaries further degrades effectiveness rather than helping.

\textbf{Summary.} Contextualized chunking is not a universal improvement. It generally benefits in-corpus retrieval (especially for LLM-guided methods), but the gains vary by model and dataset. More critically, it consistently degrades in-document retrieval. For practitioners: use contextualized chunking for in-corpus retrieval with LLM-guided segmentation, but avoid it for in-document retrieval.

\subsection{RQ4: Impact of Chunk Size}
\label{sec:rq4}

\begin{figure*}[t]
	\centering
	\includegraphics[width=\textwidth]{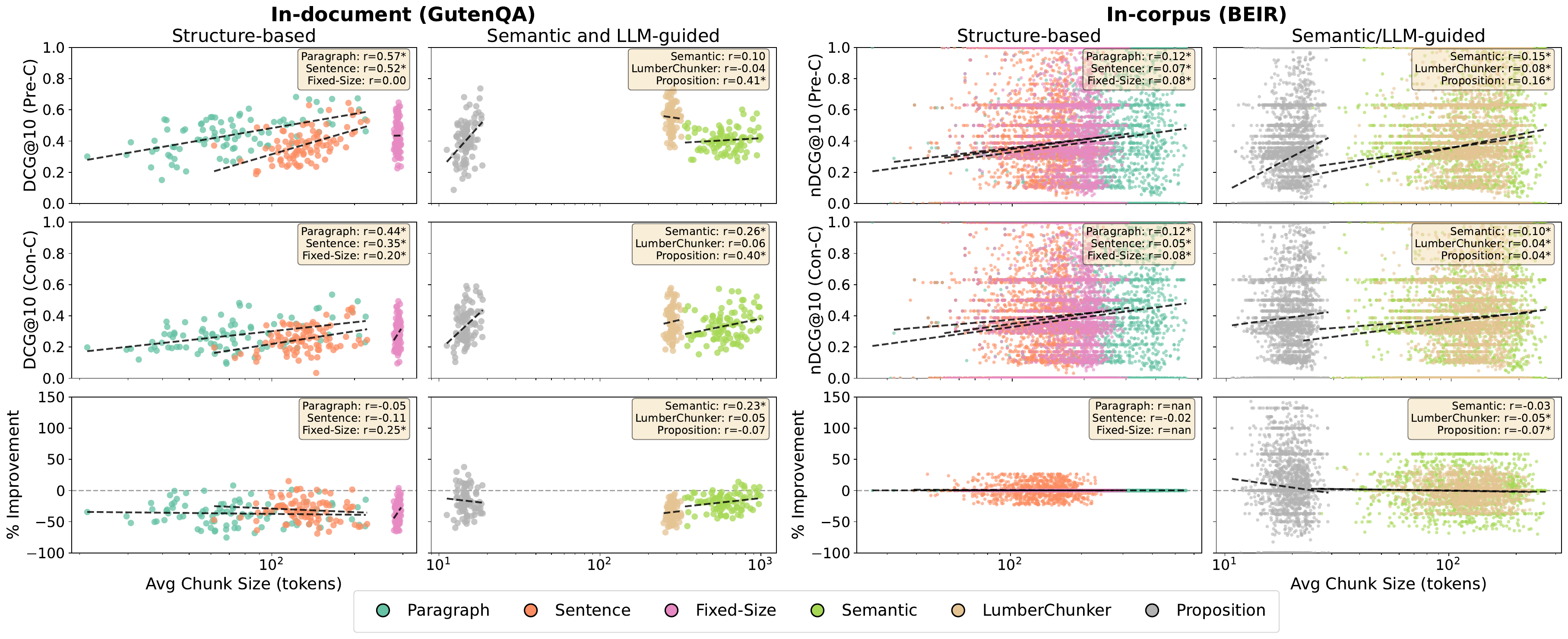}
	\caption{Per-query correlation between chunk size and retrieval effectiveness for Jina-v3. Left: in-document retrieval (GutenQA, DCG@10); Right: in-corpus retrieval (BEIR, nDCG@10). Rows: pre-embedding chunking (top), contextualized chunking (middle), relative improvement (bottom). $^*$: $p < 0.05$.}
	\label{fig:rq4}
\end{figure*}

To examine whether chunk size confounds the effectiveness differences observed across segmentation methods, we analyze the per-query relationship between chunk size and retrieval effectiveness. For each query, we compute the average chunk size in tokens\footnote{Estimated using the Jina-v3 tokenizer.} of the relevant document (for in-corpus) or relevant paragraph (for in-document) and correlate this with retrieval effectiveness. Figure~\ref{fig:rq4} presents results for Jina-v3; other models show similar patterns and are available in our repository\footnote{ \textit{Anonymoused for review}}.

\par\vspace{1em}\noindent\fcolorbox{gray!50}{gray!10}{%
	\parbox{\dimexpr\columnwidth-2\fboxsep-2\fboxrule\relax}{%
		\textbf{Finding 1:} Chunk size shows moderate correlation with in-document effectiveness but weak correlation with in-corpus effectiveness under pre-embedding chunking.
	}%
}\par\vspace{1em}

As shown in the top row of Figure~\ref{fig:rq4}, chunk size shows moderate positive correlations with in-document retrieval for paragraph-based ($r=0.57$, $p<0.05$), sentence-based ($r=0.52$, $p<0.05$), and proposition-based ($r=0.41$, $p<0.05$). In contrast, in-corpus retrieval shows weaker correlations ($r=0.08$ to $r=0.18$, though still statistically significant). One possible explanation: in in-document retrieval, larger chunks are more likely to overlap with relevant information, whereas in in-corpus retrieval, document-level aggregation (MaxP) reduces sensitivity to individual chunk sizes.

\par\vspace{1em}\noindent\fcolorbox{gray!50}{gray!10}{%
	\parbox{\dimexpr\columnwidth-2\fboxsep-2\fboxrule\relax}{%
		\textbf{Finding 2:} Contextualized chunking weakens the chunk size correlation for in-document retrieval.
	}%
}\par\vspace{1em}

Under contextualized chunking (middle row), correlations for in-document retrieval weaken: paragraph-based drops from $r=0.57$ to $r=0.44$, sentence-based from $r=0.52$ to $r=0.35$, and proposition-based from $r=0.41$ to $r=0.40$. This attenuation is consistent with contextualized chunking compensating for small chunk sizes by incorporating surrounding document context. However, correlations remain positive and mostly significant, indicating chunk size still plays a role.

\par\vspace{1em}\noindent\fcolorbox{gray!50}{gray!10}{%
	\parbox{\dimexpr\columnwidth-2\fboxsep-2\fboxrule\relax}{%
		\textbf{Finding 3:} The benefit of contextualized chunking does not depend strongly on chunk size.
	}%
}\par\vspace{1em}

The bottom row shows correlations between chunk size and percentage improvement from contextualized chunking. For in-corpus retrieval, all correlations are near zero ($r=-0.09$ to $r=0.03$), indicating consistent benefit (or harm) across chunk sizes. For in-document retrieval, patterns are mixed: sentence-based shows negative correlation ($r=-0.11$), while fixed-size shows positive correlation ($r=0.23$, $p<0.05$). The reasons for these method-specific patterns are unclear.

\textbf{Summary.} Chunk size shows stronger relationship with in-document than in-corpus retrieval effectiveness. Contextualized chunking partially reduces this relationship. For in-corpus retrieval, weak correlations suggest effectiveness differences in RQ1 are not purely driven by chunk size, though it may still play a minor role.
\section{Conclusions}

This study reproduces and examines document chunking strategies for dense retrieval, evaluating segmentation methods and embedding paradigms across in-corpus and in-document retrieval tasks.

Our findings show that optimal chunking strategies are task-dependent. For in-corpus retrieval, structure-based methods (paragraph, fixed-size) consistently outperform LLM-guided alternatives. For in-document retrieval, the pattern reverses: LumberChunker performs best, particularly on well-structured documents. Contextualized chunking improves in-corpus retrieval for LLM-guided methods but degrades in-document retrieval across all configurations. We attribute this to contextualized chunking encoding broader document themes rather than chunk-specific details, reducing discriminability between chunks within the same document.

For practitioners: use structure-based methods for in-corpus retrieval; use LumberChunker for in-document retrieval; apply contextualized chunking only for in-corpus tasks. To facilitate future research, we release our implementation at  \textit{Anonymoused for review}.


\balance
\bibliographystyle{ACM-Reference-Format}
\interlinepenalty=10000
\bibliography{bibliography.bib}

@inproceedings{rau2025context,
	author = {Rau, David and Wang, Shuai and D{\'e}jean, Herv{\'e} and Clinchant, St{\'e}phane and Kamps, Jaap},
	booktitle = {Proceedings of the Eighteenth ACM International Conference on Web Search and Data Mining},
	pages = {493--502},
	title = {Context embeddings for efficient answer generation in retrieval-augmented generation},
	year = {2025}}

@inproceedings{wang2023balanced,
	author = {Wang, Shuai and Zuccon, Guido},
	booktitle = {Proceedings of the 46th International ACM SIGIR Conference on Research and Development in Information Retrieval},
	pages = {2542--2551},
	title = {Balanced topic aware sampling for effective dense retriever: A reproducibility study},
	year = {2023}}

@inproceedings{zuccon2025r2llms,
	author = {Zuccon, Guido and Zhuang, Shengyao and Ma, Xueguang},
	booktitle = {Proceedings of the 48th International ACM SIGIR Conference on Research and Development in Information Retrieval},
	pages = {4106--4109},
	title = {R2LLMs: Retrieval and Ranking with LLMs},
	year = {2025}}

@article{li2024survey,
	author = {Li, Yongqi and Lin, Xinyu and Wang, Wenjie and Feng, Fuli and Pang, Liang and Li, Wenjie and Nie, Liqiang and He, Xiangnan and Chua, Tat-Seng},
	journal = {arXiv preprint arXiv:2404.16924},
	title = {A survey of generative search and recommendation in the era of large language models},
	year = {2024}}

@inproceedings{dai2019deeper,
	author = {Dai, Zhuyun and Callan, Jamie},
	booktitle = {Proceedings of the 42nd international ACM SIGIR conference on research and development in information retrieval},
	pages = {985--988},
	title = {Deeper text understanding for IR with contextual neural language modeling},
	year = {2019}}

@article{wang2024multilingual,
	author = {Wang, Liang and Yang, Nan and Huang, Xiaolong and Yang, Linjun and Majumder, Rangan and Wei, Furu},
	journal = {arXiv preprint arXiv:2402.05672},
	title = {Multilingual e5 text embeddings: A technical report},
	year = {2024}}

@article{nussbaum2024nomic,
	author = {Nussbaum, Zach and Morris, John X and Duderstadt, Brandon and Mulyar, Andriy},
	journal = {arXiv preprint arXiv:2402.01613},
	title = {Nomic embed: Training a reproducible long context text embedder},
	year = {2024}}

@inproceedings{thakur2021beir,
	author = {Nandan Thakur and Nils Reimers and Andreas R{\"u}ckl{\'e} and Abhishek Srivastava and Iryna Gurevych},
	booktitle = {Thirty-fifth Conference on Neural Information Processing Systems Datasets and Benchmarks Track (Round 2)},
	title = {{BEIR}: A Heterogeneous Benchmark for Zero-shot Evaluation of Information Retrieval Models},
	url = {https://openreview.net/forum?id=wCu6T5xFjeJ},
	year = {2021}}

@misc{kamradt5levels,
	author = {Kamradt, Greg},
	title = {Levels of Text Splitting},
	year = {5}}

@article{gunther2023jina,
	author = {G{\"u}nther, Michael and Milliken, Louis and Geuter, Jonathan and Mastrapas, Georgios and Wang, Bo and Xiao, Han},
	journal = {arXiv preprint arXiv:2307.11224},
	title = {Jina embeddings: A novel set of high-performance sentence embedding models},
	year = {2023}}

@article{gunther2024late,
	author = {G{\"u}nther, Michael and Mohr, Isabelle and Williams, Daniel James and Wang, Bo and Xiao, Han},
	journal = {arXiv preprint arXiv:2409.04701},
	title = {Late chunking: contextual chunk embeddings using long-context embedding models},
	year = {2024}}

@inproceedings{duarte2024lumberchunker,
	author = {Duarte, Andr{\'e} V and Marques, Jo{\~a}o DS and Gra{\c{c}}a, Miguel and Freire, Miguel and Li, Lei and Oliveira, Arlindo L},
	booktitle = {Findings of the Association for Computational Linguistics: EMNLP 2024},
	pages = {6473--6486},
	title = {Lumberchunker: Long-form narrative document segmentation},
	year = {2024}}

@inproceedings{chen2024dense,
	author = {Chen, Tong and Wang, Hongwei and Chen, Sihao and Yu, Wenhao and Ma, Kaixin and Zhao, Xinran and Zhang, Hongming and Yu, Dong},
	booktitle = {Proceedings of the 2024 Conference on Empirical Methods in Natural Language Processing},
	pages = {15159--15177},
	title = {Dense x retrieval: What retrieval granularity should we use?},
	year = {2024}}

@inproceedings{karpukhin2020dense,
	author = {Karpukhin, Vladimir and Oguz, Barlas and Min, Sewon and Lewis, Patrick SH and Wu, Ledell and Edunov, Sergey and Chen, Danqi and Yih, Wen-tau},
	booktitle = {EMNLP (1)},
	pages = {6769--6781},
	title = {Dense Passage Retrieval for Open-Domain Question Answering.},
	year = {2020}}

\end{document}